\documentclass{article}
\usepackage{spconf,amsmath,epsfig}

\usepackage{amssymb}

\title{Regularization of Building Boundaries in Satellite Images Using Adversarial and Regularized Losses}
\name{Stefano Zorzi and
        Friedrich Fraundorfer}
\address{Institute of Computer Graphics and Vision, Graz University of Technology}
\begin{document}
%
\maketitle
\begin{abstract}
In this paper we present a method for building boundary refinement and regularization in satellite images using a fully convolutional neural network trained with a combination of adversarial and regularized losses.
Compared to a pure Mask R-CNN model, the overall algorithm can achieve equivalent performance in terms of accuracy and completeness.
However, unlike Mask R-CNN that produces irregular footprints, our framework generates regularized and visually pleasing building boundaries which are beneficial in many applications.
\end{abstract}


\section{Introduction}
\label{sec:intro}
Building detection and segmentation from satellite images is still a challenging problem.
Automatically detecting constructions and extracting precisely their footprints is in the interest of many engineering and cartographic applications.
In recent years, multiple machine learning challenges have been proposed to encourage people to present new building extraction methods (e.g. Deep Globe Challenge\footnote{http://deepglobe.org/challenge.html}, SpaceNet Challenge\footnote{https://spacenetchallenge.github.io/}, CrowdAI Mapping Challenge\footnote{https://www.crowdai.org/challenges/mapping-challenge}).

The most common and effective way to deal with this problem is the use of powerful semantic segmentation or instance segmentation networks.
However, in most cases, predicted building footprints have irregular shapes which are very different from the ones used in cartographic applications.

This problem has been recently dealt with Kang et al.~\cite{zhao2018buildingRegularization} where they proposed a building segmentation and refinement pipeline as a solution for the DeepGlobeChallenge 2018. 
Their framework is composed of a Mask R-CNN~\cite{he2017mask} model for instance segmentation followed by a boundary refinement algorithm that exploits polygon simplification methods.
The overall algorithm produces more realistic building footprints, but it does not consider the intensity image for the regularization to further improve the results.

In this paper we present a new building segmentation and regularization framework completely based on Deep Learning techniques.
The pipeline itself is the same as Kang's, so we still perform the building segmentation as a first step and then we apply the building regularization as a second step.
The difference is in the use of a fully convolutional neural network as a regularization method, instead of using polygon simplification algorithms.

Inspired by deep style transfer techniques like pix2pix~\cite{pix2pix} and cyclegan~\cite{zhu2017cyclegan}, we train our regularization network using adversarial losses to produce more realistic footprints.
In particular, we use OpenStreetMap building footprints as the target footprint domain to train a GAN~\cite{goodfellow2014generative} architecture. 
We also exploit regularized losses~\cite{tang2018regularizedLosses,tang2018ncutLoss} to make the network aware of the real building boundaries in the intensity image and, consequently, to further refine the result.
Finally, a reconstruction loss ensures to obtain regularized footprints that look similar in size, pose and shape to the original Mask R-CNN predicted footprints.

The combination of these three types of loss functions enables us to learn a regularization network that not only produces better looking and more realistic building footprints, but is also capable of achieving better scores on the test dataset compared to the pure Mask R-CNN solution.

\section{Method}
\label{sec:formulation}
\begin{figure*}[thbp]
\centering
\includegraphics[width=1\linewidth]{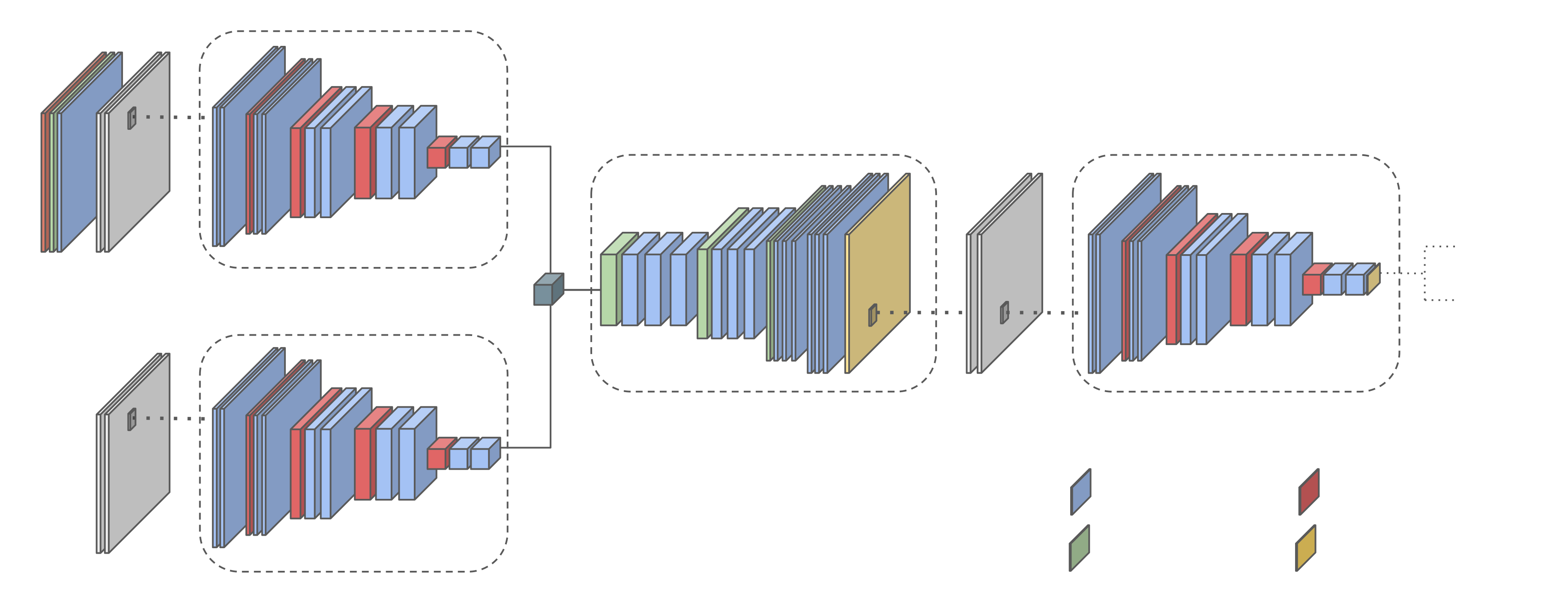}
\put (-75,20) {\scriptsize{conv 1$\times$1}, sigmoid}
\put (-75,37) {\scriptsize{max pool 2$\times$2}}
\put (-150,42) {\scriptsize{conv 3$\times$3},}
\put (-150,34) {\scriptsize{batch norm, ReLU}}
\put (-150,20) {\scriptsize{max pool 2$\times$2}}
\put (-210,65) {\scriptsize{Either regularized or}}
\put (-208,57) {\scriptsize{reconstructed mask}}
\put (-375,100) {\scriptsize{Latent Space}}
\put (-500,100) {\scriptsize{Image}}
\put (-475,100) {\scriptsize{Input mask}}
\put (-485,5) {\scriptsize{Ideal mask}}
\put (-475,80) {\textbf{$y$}}
\put (-490,185) {\textbf{$z$}}
\put (-465,185) {\textbf{$x$}}
\put (-365,175) {\textbf{$E_G$}}
\put (-365,75) {\textbf{$E_R$}}
\put (-305,135) {\textbf{$F$}}
\put (-75,135) {\textbf{$D$}}
\put (-34,117) {\scriptsize{$true$}}
\put (-34,98) {\scriptsize{$false$}}
\caption{Workflow of the proposed regularization framework. It is composed of two paths: the generator path ($E_G \rightarrow F$) produces the regularized building footprint mask; the reconstruction path ($E_R \rightarrow F$) encodes and decodes the ideal input mask ensuring to have the same real valued masks as input to the discriminator.}
\label{fig:workflow}
\end{figure*}
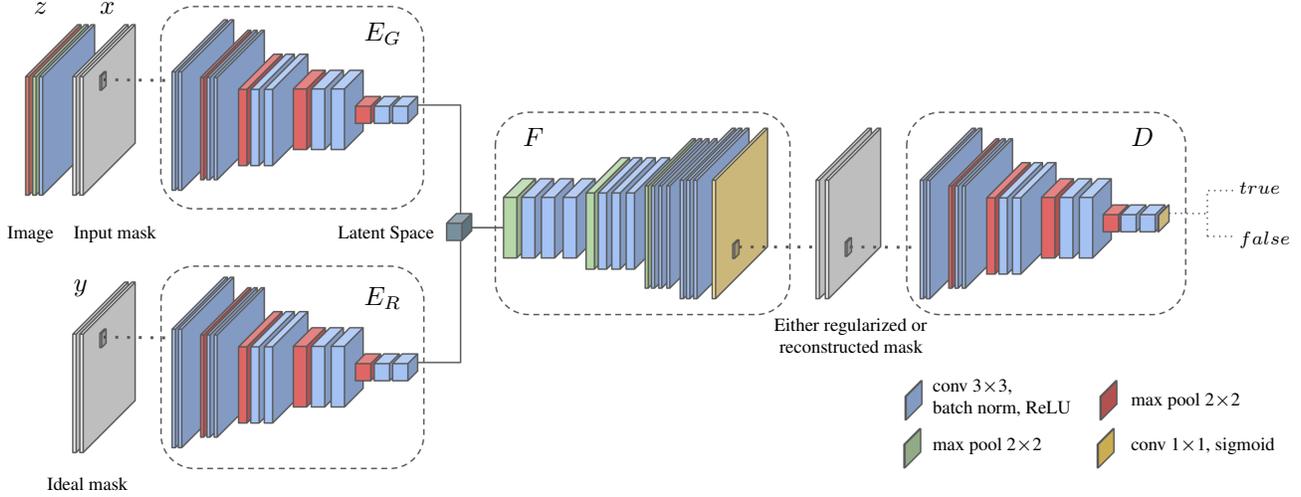


Our aim is to learn a mapping function between the domain $X$ (Mask R-CNN footprints) and the domain $Y$ (ideal footprints) given the training samples $\{x_i\}^{N}_{i=1}$ where $x_i \in X$ and $\{y_i\}^{M}_{i=1}$ where $y_i \in Y$.
We also exploit RGB images, $\{z_i\}^{N}_{i=1}$ where $z_i \in Z$, to further improve the results training the model with an additional regularized loss.

The model performs the regularization $G:\{X,Z\} \rightarrow Y$ exploiting an encoder-decoder network, as shown in Figure~\ref{fig:workflow}.
The generation of the regularized footprints is performed by the encoder $E_G$ and the decoder $F$, so $G$ can be seen as the combination of the two: $G(x,z) = F(E_G(x,z))$.
A discriminator $D$ is introduced in order to distinguish between regularized footprints $G(x,z)$ and ideal ones.
It is worth noting that the ideal building footprints are not directly evaluated by the discriminator model, but the ideal mask is encoded by $E_R$ and decoded back by the common network $F$. 
The aim of this path is to obtain a reconstructed version of $y$.
One concern for this design choice is that the adversarial network can potentially trivially distinguish the two distributions by detecting if the mask consists of zeros and ones (one-hot encoding of the ideal mask), or of real values between zero and one (output of the autoencoder). 
This problem is solved by generating both reconstructed and regularized samples with the same network $F$.
Also, this architecture ensures stability during training and avoids a winning discriminator situation since the two autoencoders are connected (with the common decoder) and trained together.

The encoders and the decoder are learned exploiting three types of loss functions: 
\textit{adversarial loss}, \textit{reconstruction losses} and \textit{regularized loss}.

\subsection{Adversarial Loss}
We use adversarial losses~\cite{goodfellow2014generative} to learn the mapping function between the domain $X$ and $Y$. 

The objective function used to learn the discriminator $D$ is expressed as:
\begin{equation}
\label{eqation:loss_gan_D}
\begin{split}
\mathcal{L}_{D}(G,R,D) &= \mathbb{E}_{y}[(1-D(R(y)))^2] 
\\&+ \mathbb{E}_{x,z}[D(G(x,z))^2]
\end{split}
\end{equation}

where the path $R(y)=F(E_R(y))$ encodes and reconstructs the ideal mask and the path $G(x,z)=F(E_G(x,z))$ generates building footprints that look similar to ideal footprints in domain $Y$.
The aim of $D$ is to distinguish between regularized footprints and reconstructed footprints.
Note that we used the least-squared loss in equation \ref{eqation:loss_gan_D} because it ensures better stability during training and generates higher quality results~\cite{zhu2017cyclegan}.

For the mapping path $G$ the loss function is expressed as:
\begin{equation}
\label{eqation:loss_gan_G}
\begin{split}
\mathcal{L}_{GAN}(G,D) = \mathbb{E}_{x,z}[(1-D(G(x,z))^2]
\end{split}
\end{equation}

This path is trained to fool the discriminator $D$, in fact, the adversarial loss encourages $G$ to produce footprints similar to the samples on the $Y$ domain.

\subsection{Reconstruction Loss}
In order to force the network to generate building footprints similar to the input masks, we simply use the \textit{binary cross entropy loss} both on the generator path $G$ and on the reconstruction path $R$.
The loss is computed between $x$ and $G(x)$ and between $y$ and $R(y)$ to produce regularization masks close to the Mask R-CNN predictions and to the ideal masks, respectively.
The two losses can be expressed as:
\begin{equation}
\begin{split}
\mathcal{L}_{BCE_G}(G) &= -\sum_{i}^{N} x_i \cdot \log G(x,z)_i
\\
\mathcal{L}_{BCE_R}(R) &= -\sum_{i}^{N} y_i \cdot \log R(y)_i
\end{split}
\end{equation}


\subsection{Regularized Loss}
Without \textit{regularized losses} our model would not be able to exploit image information to further improve the building regularization.

Alongside the adversarial loss and the reconstruction loss, the \textit{Potts loss}~\cite{tang2018regularizedLosses} and the \textit{normalized cut loss}~\cite{tang2018regularizedLosses,tang2018ncutLoss} are used to learn our model.
These two loss functions force the generator $G$ to produce building footprints aligned to the building boundaries observed in the intensity image.
Also, trained with these losses, the generator is capable of solving some artifacts produced by Mask R-CNN (Figure \ref{fig:examples}). 

Potts and normalized cut loss functions can be expressed as:
\begin{equation}
\label{eq:potts}
\mathcal{L}_{potts}(G) = \sum_{k}^{} S^{k\top} W (1-S^k)
\end{equation}
\begin{equation}
\label{eq:ncut}
\mathcal{L}_{ncut}(G) = \sum_{k}^{} \frac{S^{k\top} \hat{W} (1-S^k)}{1^\top \hat{W} S^k}
\end{equation}

where $W$ and $\hat{W}$ are a matrices of pairwise discontinuity costs or \textit{affinity matrices}, while $S = G(x,z)$ is the k-way softmax segmentation mask generated by the network.
$S^k$ describes the vectorization of the $k$-th channel in the segmentation image. 
In our case $k=2$ since we have two classes.

\subsection{Full Objective}
The full objective used to train the generator $G$ and the reconstruction $R$ model is a linear combination between the \textit{adversarial loss}, the \textit{reconstruction loss} and the \textit{regularized loss}.
\begin{equation}
\label{eq:full_objective}
\begin{split}
\mathcal{L}_{}(G,R,D) &= \alpha \mathcal{L}_{GAN}(G,R,D)\\
 &+ \beta \mathcal{L}_{BCE_G}(G) + \gamma \mathcal{L}_{BCE_R}(R) \\
 &+ \delta \mathcal{L}_{Potts}(G) + \epsilon \mathcal{L}_{ncut}(G)
\end{split}
\end{equation}

Note that the losses through the paths $G$ and $R$ are obtained switching the encoders $E_G$ and $E_R$. 
Once the total loss has been computed, the backpropagation step is performed and the weights of $E_G$, $E_R$ and $F$ are updated jointly.

\section{Implementation Details}
\subsection{Dataset}
We trained our regularization framework on a satellite image which represents the city of Jacksonville, Florida. 
The image is obtained by performing the pansharpening between the panchromatic layer and three multispectral channels (infrared, green, blue). 
There is no technical reason why we use the infrared channel. 
The decision has been taken just for a visualization preference, since grass and trees highlighted in red make the roofs of the buildings more visible to the naked eye.
Input masks are generated by a Mask R-CNN model trained using OpenStreetMap footprints.
OpenStreetMap footprints are also used as ideal masks during the regularization framework training.
In order to achieve better results, our models are learned using single building instances instead of patches.
As a test-set, we manually labeled an image of a residential area in Jacksonville mainly composed of mid-sized and small-sized buildings.
The size of the test area is around 360$\times$620 squared meters and it counts 243 buildings.


\subsection{Network Architecture}
The network follows the same design choices of a classical convolutional autoencoder, as shown in Figure \ref{fig:workflow}.
The encoders $E_G$, $E_R$ and the discriminator $D$ share the same architectural design.
They are composed of a chain of 3$\times$3 convolutional layers with stride 1, followed by batch normalization layers and 2$\times$2 pooling layers.
After every down-sampling operation, height and width of the tensor are halved, but the number of convolutional filters is doubled.

The decoder network $F$ has the dual architecture. 
It is composed of a chain of 3$\times$3 convolutional, batch normalization and up-sampling layers.
This time, after every up-sampling layer, the resolution increases but the number of channels of the tensor is gradually reduced.


\subsection{Training Details}
For the training every building mask and the corresponding RGB picture are resized to 256$\times$256 pixels images.
The ideal masks are generated drawing the OpenStreetMap building footprint polygons in 256$\times$256 pixels masks as well.

For all the experiments we use Adam optimizer with a batch size of 8. The models are trained for 80000 batches in total.
All networks are learned from scratch with an initial learning rate of 0.0002.
We keep the same learning rates for 60000 batches and linearly decay the rates to zero over the last 20000 batches.

We set $\alpha=3$, $\beta=3$, $\gamma=1$, $\delta=200$ and $\epsilon=2$ in Equation \ref{eq:full_objective}. 
$\epsilon$ and $\delta$ are linearly increased from zero to 2 and 200, respectively, during the first 30000 batches to keep the learning more stable.

The weight matrix $W$ and $\hat{W}$ for \textit{potts loss} and \textit{normalized cut loss} are constructed as:
\begin{equation}
\label{eq:weights}
w_{ij} = e^{\frac{-\|F(i)-F(j)\|_2^2}{\sigma_I^2}} \cdot     
\begin{cases}
      e^{\frac{-\|X(i)-X(j)\|_2^2}{\sigma_X^2}} & {\scriptstyle \text{if}\ \|X(i)-X(j)\|_2 < r} \\
      0 & {\scriptstyle \text{otherwise} }
\end{cases}
\end{equation}

where $X(i)$ and $F(i)$ are the spatial location and pixel value of node $i$, respectively. In Equation \ref{eq:weights} we use $\sigma_I = 0.075$, $\sigma_X = 4$ and $r=19$ both for $W$ and $\hat{W}$.
Images and masks have values normalized between 0 and 1.

\section{Experimental Results}
\begin{table}[]
\begin{tabular}{|c|c|c|c|c|}
\hline
Metric              & Recall         & Precision      & $F_{0.5}$         & IoU  \\ \hline
Mask R-CNN           & 0.885          & \textbf{0.933} & 0.923 		& 0.833          \\
Our (no reg. loss) & 0.854          & 0.932          & 0.916          & 0.805          \\
Our                & \textbf{0.909} & 0.932          & \textbf{0.927} & \textbf{0.852} \\ \hline
\end{tabular}
\caption{Scores of building extraction computed on the test area.}
\label{table:results}
\end{table}

The performances of our algorithm are evaluated based on the Intersection over Union (IoU) metric.
Computing the scores, we want to analyze the effects of building regularization on the building extraction, comparing the result of the pure Mask R-CNN model with the result of the regularization pipeline (Mask R-CNN and regularization).

Table~\ref{table:results} shows the scores for Mask R-CNN and our method.
Although Mask R-CNN shows slightly higher precision values, our regularization pipeline achieves higher results on recall, $F_{0.5}$ and Jaccard index (IoU) scores.

We also trained a model without Potts and normalized cut losses. 
The scores show higher results for the complete regularization method, a sign that the regularized losses are effective and can be used to refine the segmentation results.

\begin{figure}
\centering
\begin{tabular}{cc}
  \includegraphics[width=32mm]{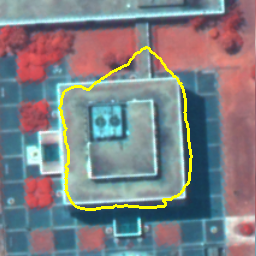} &   \includegraphics[width=32mm]{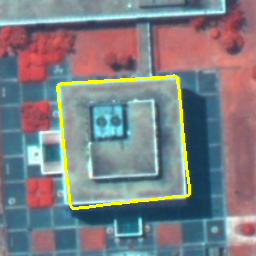} \\
 \includegraphics[width=32mm]{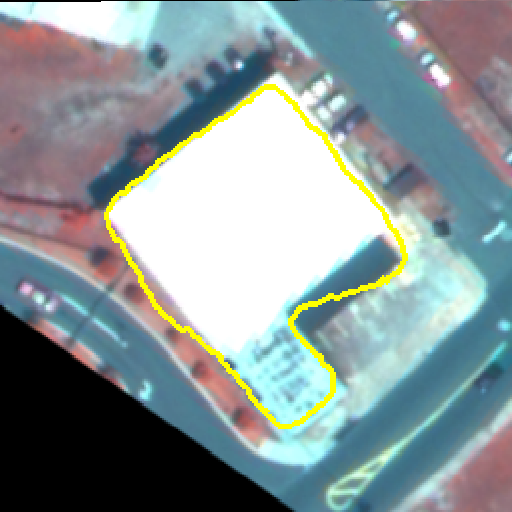} &   \includegraphics[width=32mm]{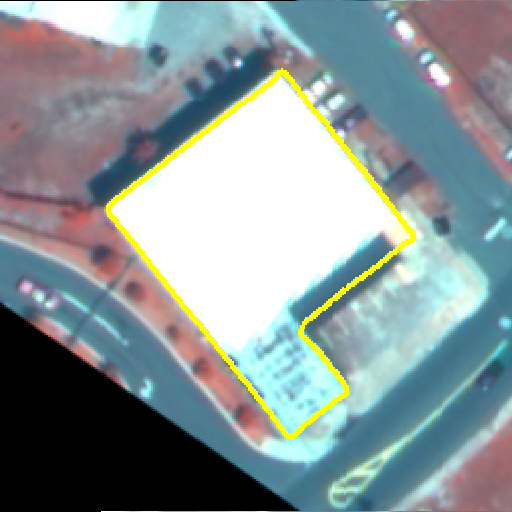} \\
  \includegraphics[width=32mm]{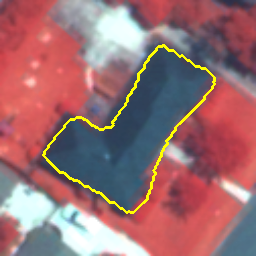} &   \includegraphics[width=32mm]{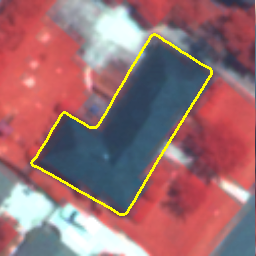} \\
\end{tabular}
\caption{Comparison between footprints produced by Mask R-CNN (left column) and our regularization method (right column).}
\label{fig:examples}
\end{figure}

\begin{figure}
\centering
\begin{tabular}{cc}
  \includegraphics[width=40mm]{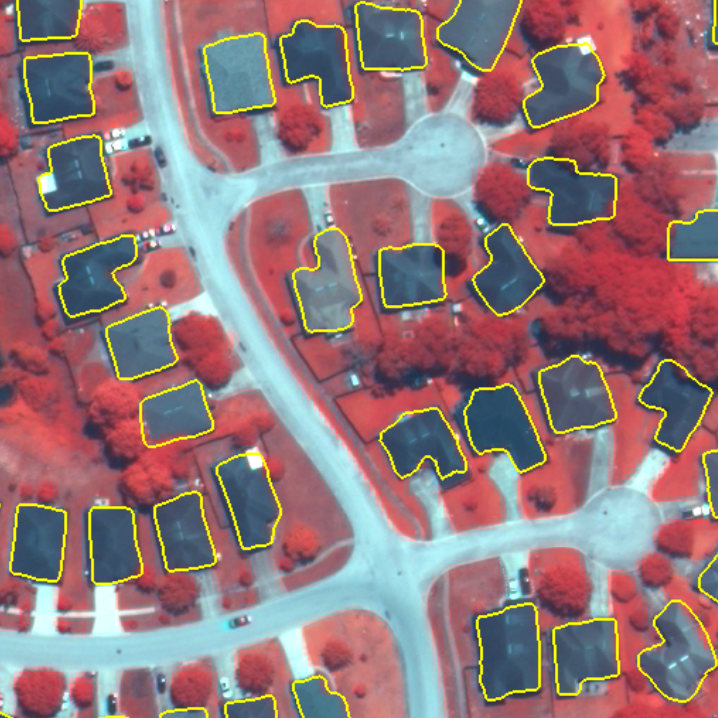} &   \includegraphics[width=40mm]{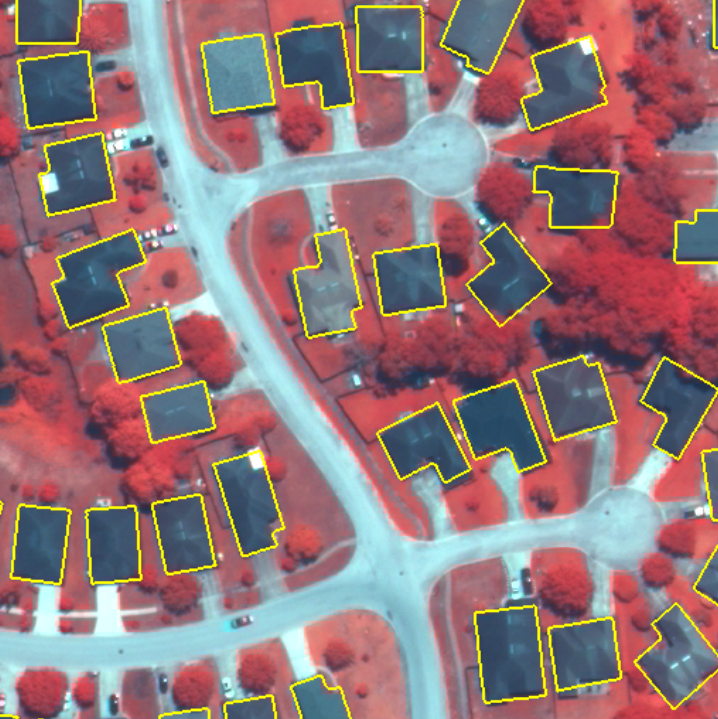} \\
\end{tabular}
\caption{Portion of the test area evaluated by Mask R-CNN (left) and regularized by our framework (right).}
\label{fig:test}
\end{figure}

To summarize, our method produces better representations of building footprints with more regular boundaries. 
Some regularization examples are shown in Figure~\ref{fig:examples}, while a regularized portion of the test area is shown in Figure~\ref{fig:test}.

\section{Conclusions}

We presented a building extraction method that combines a Mask R-CNN model for instance segmentation with a network for footprints regularization. 
The regularization network has proved capable of exploiting effectively the information of the intensity image to further refine building boundaries, achieving equivalent or even higher results in terms of Intersection over Union compared to the pure Mask R-CNN model.
Moreover, unlike Mask R-CNN that produces irregular building masks, our method generates regularized footprints that can be used in many cartographic and engineering applications.

\bibliographystyle{IEEEbib}
{\small \bibliography{main}}

\end{document}